\documentclass[twocolumn,prl,aps,amsfonts,amssymb,amsmath,showpacs]{revtex4-1}
\usepackage{graphicx}
\usepackage{dcolumn}
\usepackage{mathtools}
\graphicspath{{figures/}}
\setkeys{Gin}{width=\linewidth}

\begin{document}
\title{Noise suppression using symmetric exchange gates in spin qubits}

\author{Frederico Martins$^{1,*}$, Filip K. Malinowski$^{1,*}$, Peter D. Nissen$^{1}$, Edwin Barnes$^{2,3}$, Saeed Fallahi$^{4}$, Geoffrey C. Gardner$^{5}$, Michael J. Manfra$^{4,5,6}$, Charles M. Marcus$^{1}$, Ferdinand Kuemmeth$^{1}$\\}

\affiliation{ $^{1}$ Center for Quantum Devices, Niels Bohr Institute, University of Copenhagen, 2100 Copenhagen, Denmark\\
$^{2}$ Department of Physics, Virginia Tech, Blacksburg, Virginia 24061, USA\\
$^{3}$ Condensed Matter Theory Center and Joint Quantum Institute, Department of Physics, University of Maryland, College Park, Maryland 20742-4111, USA\\
$^{4}$ Department of Physics and Astronomy and Birck Nanotechnology Center, Purdue University, West Lafayette, Indiana 47907, USA\\
$^{5}$ School of Materials Engineering and Birck Nanotechnology Center, Purdue University, West Lafayette, Indiana 47907, USA\\
$^{6}$ School of Electrical and Computer Engineering, Purdue University, West Lafayette, Indiana 47907, USA\\
$^{*}$ These authors contributed equally to this work}

\date{\today}

\begin{abstract}
We demonstrate a substantial improvement in the spin-exchange gate using symmetric control instead of conventional detuning in GaAs spin qubits, up to a factor-of-six increase in the quality factor of the gate. For symmetric operation, nanosecond voltage pulses are applied to the barrier that controls the interdot potential between quantum dots, modulating the exchange interaction while maintaining symmetry between the dots. Excellent agreement is found with a model that separately includes electrical and nuclear noise sources for both detuning and symmetric gating schemes. Unlike exchange control via detuning, the decoherence of symmetric exchange rotations is dominated by rotation-axis fluctuations due to nuclear field noise rather than direct exchange noise. 
\end{abstract}

\maketitle

Spin qubits, basic units of quantum information built from the spin states of electrons in solid-state systems, are one of the most promising realizations of a qubit~\cite{Kloeffel_2013}. This is due to their potential for minituarization, scalability and fault tolerance~\cite{Taylor_2005, Awschalom_2013}. In fact, experiments in recent years have demonstrated remarkable progress in the coherent manipulation of single- and multi-spin devices~\cite{FolettiNatPhys2009,Bluhm_2010_1,Petta_2010,ShulmanScience2012}.
Nevertheless, one of the main difficulties with spin qubits, and more generally with solid-state qubits, is the decoherence due to interactions with the environment.
In the case of electron spins confined in semiconductor quantum dots, two main types of environmental noise limit coherence: electrical noise and hyperfine interactions with nuclear spins in the surrounding lattice~\cite{Bluhm_2010_2,Dial_2010,Barthel_2012}. 
To reach the high control fidelities necessary for quantum computing, the coupling between a quantum dot spin qubit and its environment can be reduced by the use of sweet spots~\cite{WongPRB2015, HiltunenPRB2015, FeiPRB2015}, and pulse errors can be reduced by bootstrap tomography~\cite{DobrovitskiPRL2010, CerfontainePRL2014}.

\begin{figure}
\begin{center}
\includegraphics[width=86 mm]{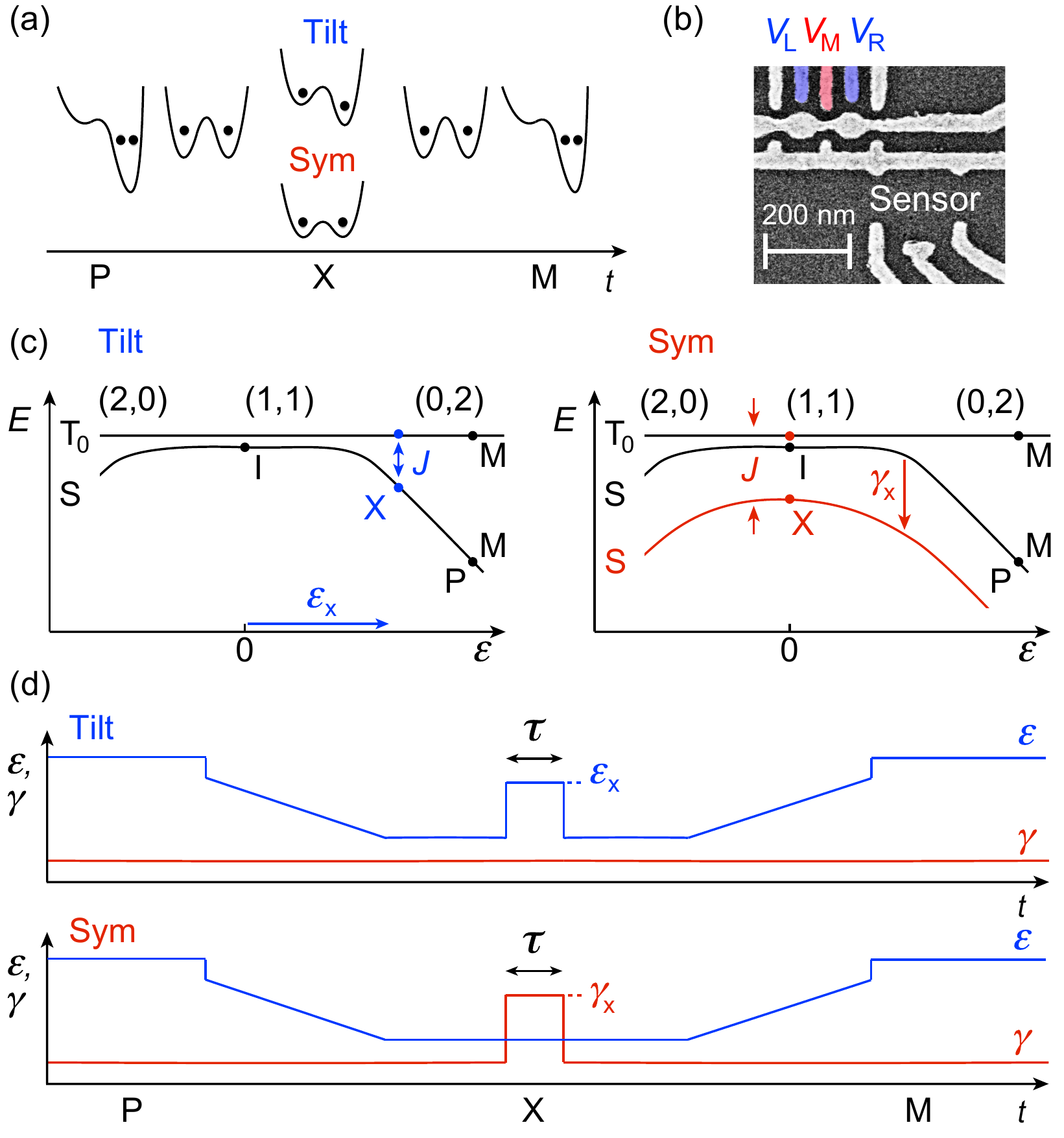}
\caption{(Color online)
(a) Schematic comparison of detuning (tilt) and symmetric exchange pulse sequences, showing double-dot potentials and dot occupancies.
Tilt: wave function overlap controlled by detuning the confining potential; Symmetric: wave function overlap controlled by lowering the potential barrier between dots.
(b) Electron micrograph of the device consisting of a double dot and charge sensor. Note the gate that runs through the center of the dots. A 10 nm HfO$_2$ layer is deposited below the gates to allow positive and negative gating. 
High-bandwidth lines are connected to left and right plungers gates, $V_{\mathrm{L}}$, $V_{\mathrm{R}}$ (blue), and the middle barrier gate,  $V_{\mathrm{M}}$ (red).
(c) Energy diagrams of the two-electron spin singlet, S, and spin-zero triplet, T$_0$,  as a function of detuning $\varepsilon$.
(Left) Tilt mode:  Exchange, $J$, is controlled by detuning $\varepsilon$, set by $V_{\mathrm{L}}$ and $V_{\mathrm{R}}$; (Right) Symmetric mode: $J$ is controlled interdot coupling, $\gamma$, set by $V_{\mathrm{M}}$ (red curve).
(d) Pulse sequences for tilt and symmetric modes, with amplitudes $\varepsilon_{\mathrm{x}}$ and  $\gamma_{\mathrm{x}}$ during the exchange pulse, respectively.
}
\label{fig:fig1}
\end{center}
\end{figure}

A crucial component of any spin-based quantum computing platform is strong spin-spin interaction.
In their seminal article, Loss and DiVincenzo proposed that exchange interactions between electron spins could be controlled by the height of the tunnel barrier between neighboring quantum dots~\cite{Loss_1998}.
However, until recently this proposal was not implemented in the laboratory, and instead exchange interactions were induced by raising or lowering the potential of one dot relative to the other, an approach referred to as tilt or detuning control~\cite{Petta_2005}.  Unlike the dot-symmetric tunnel barrier control method, tilt control affects the two dots asymmetrically and hybridizes the (1,1) and (0,2) charge states. 
Here numbers within each parenthesis denote occupation number of the left dot and right dot. 
In Fig.~\ref{fig:fig1}(a) we illustrate the difference between the two methods. Firstly, a singlet state (0,2)S is prepared (P). 
Thereafter the electrons are adiabatically separated to the $|$${\uparrow\downarrow\rangle}$ state in the (1,1) charge configuration.  At the exchange point (X), a pulse is performed. For the tilt case, during this pulse the wavefunctions of the electrons are brought together by asymmetrically deforming the confining potential of the dots. In the case of the symmetric mode of operation, the exchange interaction is increased by lowering the potential barrier between the two dots. Finally, reversing the slow adiabatic passage first projects the final two-spin state onto $|$${\uparrow\downarrow\rangle}$ and then maps it onto (0,2)S, which is then read out at the measurement point (M).

In this Letter, we demonstrate rapid, high-quality exchange oscillations implemented by pulsing the barrier between two dots, as envisioned in the original Loss-DiVincenzo proposal. 
We also show that, unlike tilt-induced qubit rotations, the coherence of barrier-induced rotations is not limited by electrical detuning noise, but rather by nuclear spin fluctuations parallel to the applied magnetic field. 
We quantify the improvements by studying exchange oscillations within a singlet-triplet qubit, corresponding to $\sqrt{\mathrm{SWAP}}$ operations between the two spins. 
Alternatively benchmarking of single-qubit gate fidelities is in principle possible but requires nuclear programming~\cite{FolettiNatPhys2009}. 
Recent work on surface acoustic waves and silicon triple quantum dots showed results consistent with some of our observations~\cite{Bertrand_2014, Reed_2015}, indicating that symmetric exchange finds applications beyond GaAs qubits.  

The double quantum dot device with integrated charge sensor \cite{Barthel_2009} is shown in Fig.~\ref{fig:fig1}(b). 
The device was fabricated on a GaAs/AlGaAs heterostructure 57~nm below the surface, producing a two-dimensional electron gas with bulk  density $n$~=~$2.5\times10^{15}$~m$^{-2}$ and mobility  $\mu$~=~230~m$^{2}$/Vs. 
To minimize stray capacitance a mesa was patterned using electron-beam lithography and wet etching. Metallic gates (Ti/Au) were deposited after atomic layer deposition of 10~nm HfO$_2$, which allows both positive and negative gating, and obviates gate-bias cooling~\cite{Buizert_2008}.
 All measurements were conducted in a dilution refrigerator with mixing chamber temperature below 50 mK  and in-plane magnetic field $B = 300$~mT applied perpendicular to the axis between dots.

Voltages pulses were applied via high-bandwidth coaxial lines to the left and right plunger gates, $V_{\mathrm{L}}$, $V_{\mathrm{R}}$, and the barrier between the dots, $V_{\mathrm{M}}$. In practice, to account for the small coupling assymmetries, all three gates are involved in applying detuning $\varepsilon$ and symmetric barrier control $\gamma$:

\begin{align}
	\label{definition}
	\begin{split}
	\varepsilon &=k_{\mathrm{0}}[(V_{\mathrm{R}}-V_{\mathrm{R}}^0)-(V_{\mathrm{L}}-V_{\mathrm{L}}^0)]+k_{\mathrm{1}}(V_{\mathrm{M}}-V_{\mathrm{M}}^0),
	\\
	\gamma &=V_{\mathrm{M}}-V_{\mathrm{M}}^0, 
	\end{split}
\end{align}
where $V_{\mathrm{R}}^0$, $V_{\mathrm{L}}^0$ and  $V_{\mathrm{M}}^0$ are DC offset voltages (see Supplementary Material). Parameters $k_{\mathrm{0}} = 0.5$ and $k_{\mathrm{1}}=-0.075$ were determined  experimentally by mapping out the charge stability diagram. The value of $k_{\mathrm{0}}$ is consistent with previous experiments and sets the difference between left and right dot electrochemical potential, whereas $k_{\mathrm{1}}$ keeps other charge states  energetically unaccessible during $\gamma$ pulses.

Energy levels for the two-electron singlet S and triplet T$_0$ states as a function of detuning, $\varepsilon$,  are shown in Figs.~1(c), along with the pulse sequences for the tilt and symmetric operation modes in Fig.~1(d). For both  tilt and symmetric operation, two electrons are prepared (P) in a singlet (0,2)S state and, by slowly ramping $\varepsilon$ to (1,1), the system is initialized (I) into the ground state  of the nuclear Overhauser field, either $|$${\uparrow\downarrow\rangle}$ or $|$${\downarrow\uparrow\rangle}$.  For tilt operation, the exchange pulse, $J$, is applied by detuning to the exchange (X) point $\varepsilon_{\mathrm{x}}$ for a duration $\tau$, inducing rotations between $|$${\uparrow\downarrow\rangle}$ and $|$${\downarrow\uparrow\rangle}$. For symmetric operation, the exchange pulse is applied by pulsing the middle gate to $\gamma_{\mathrm{x}}$.

\begin{figure}
\begin{center}
\includegraphics[width=86 mm]{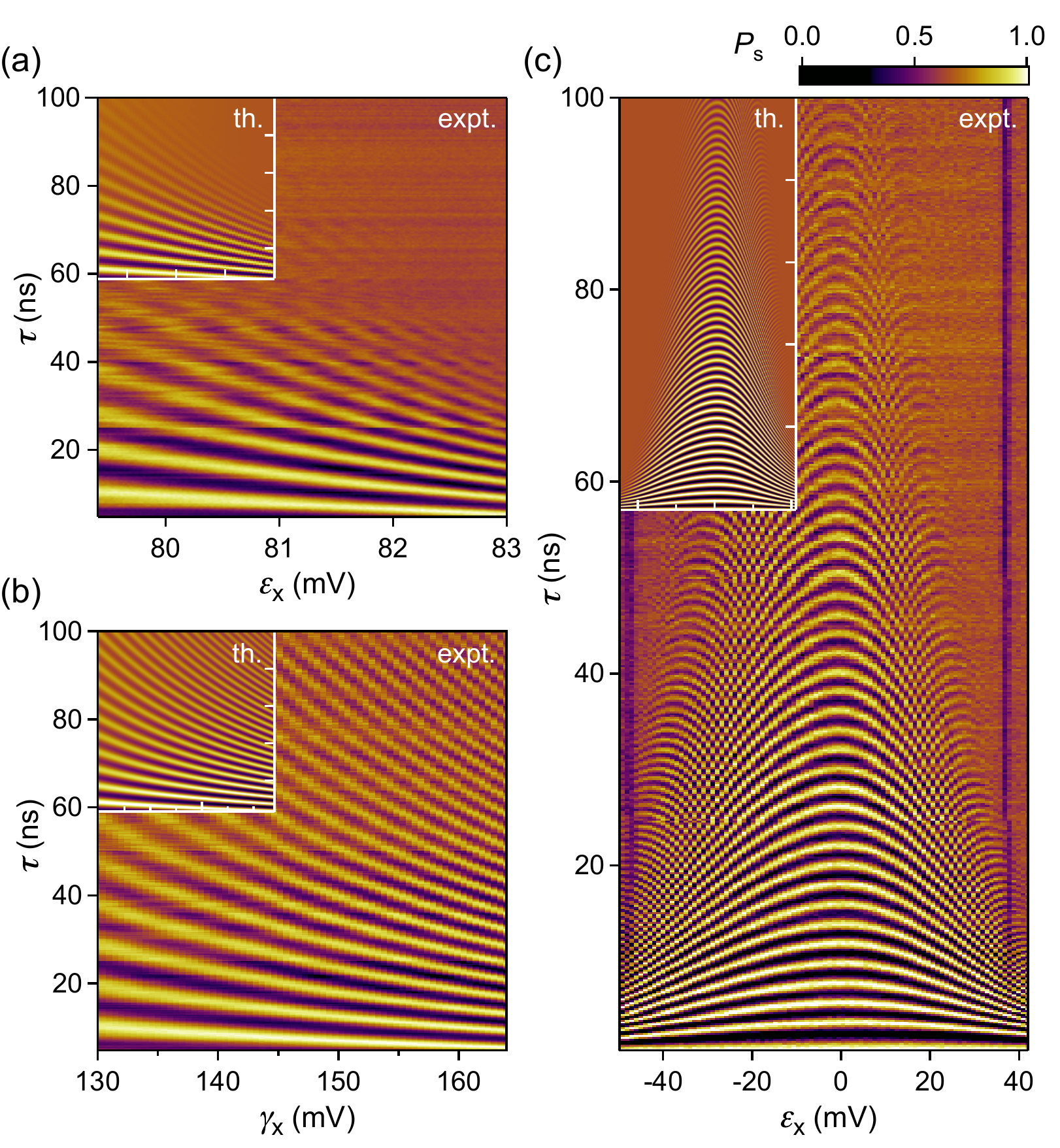}
\caption{(Color online)
(a) Probability of detecting a singlet, $P_{\mathrm{s}}$, as a function of $\varepsilon_{\mathrm{x}}$ and
exchange time $\tau$ for tilt-induced oscillations ($\gamma_{\mathrm{x}}$ = 0 mV).
(b) $P_{\mathrm{s}}$ as a function of $\gamma_{\mathrm{x}}$ and exchange time $\tau$ obtained for barrier-induced oscillations near the symmetry point ($\varepsilon_{\mathrm{x}}$ = 13.5 mV).
(c) Same as (a) with barrier pulse activated, $\gamma_{\mathrm{x}}$ = 190 mV, revealing the sweet spot of the symmetric operation. 
The dark vertical features near 39~mV and -44 mV are due to leakage from the singlet state to the spin-polarized triplet state.
Insets show theoretical simulations for each experimental situation.
}
\label{fig:fig2}
\end{center}
\end{figure}

Two-dimensional images of exchange oscillations, controlled by either tilt [Fig.~\ref{fig:fig2}(a)] or symmetric operation near the midpoint of (1,1) [Fig.~\ref{fig:fig2}(b)], show a striking difference in quality. 
In both images, each pixel represents the singlet return probability, $P_{\mathrm{S}}$, measured from an ensemble of $\sim 10^{3}$ single-shot measurements. Each single-shot measurement is assigned a binary value by comparing the reflectometer signal at the measurement (M) point, integrated for $T_\mathrm{M}=10$~$\mu$s, to a fixed threshold  \cite{Barthel_2009,Barthel_2010}. 
Figure \ref{fig:fig2}(c) shows exchange oscillations using both tilt and exchange. This image is generated by applying a tilt pulse of amplitude $\varepsilon_{\mathrm{x}}$ (of either sign) along with a fixed symmetric pulse $\gamma_{\mathrm{x}} = 190$~mV for a duration $\tau$. As $|\varepsilon_{\mathrm{x}}|$ is increased $J$ also increases, producing a chevron-like pattern centered around the sweet spot $J(\varepsilon_{\mathrm{x}} = 0)$ that occurs in the middle of the (1,1) charge state. 
Defining a quality factor, $Q$, to be the number of oscillations before the amplitude decays to $1/e$ of its initial value, we measure $Q\sim 35$ at the symmetry point, $\varepsilon_{\mathrm{x}} = 0$ ~\cite{definitionsymm}.

\begin{figure}
\begin{center}
\includegraphics[width=86 mm]{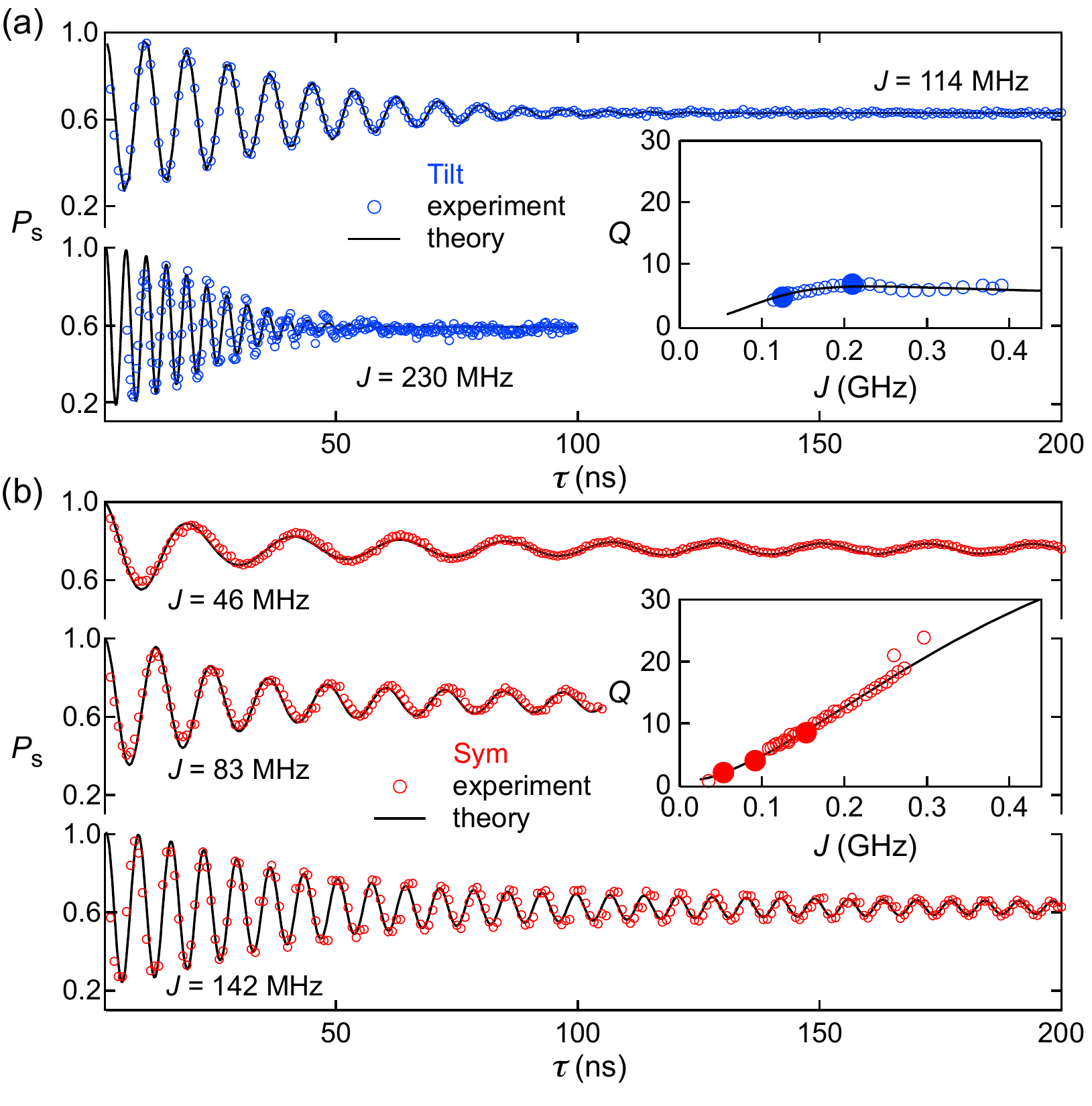}
\caption{(Color online)
(a) Tilt-induced exchange oscillations ($i.e.$ $\gamma_{\mathrm{x}}=$~0~mV) for  $\varepsilon_{\mathrm{x}} = $~79.5~mV and 82~mV, generating oscillation frequencies indicated by $J$.
(b) Same as (a) but for the symmetric mode of operation ($\varepsilon_{\mathrm{x}}=$~13.5~mV), with $\gamma_{\mathrm{x}}=$~100~mV, 120 mV and 140 mV. 
Open circles are experimental data. 
Solid lines correspond to the theoretical model in Eq.~\eqref{Ed_Model}, with $J$ and a horizontal offset being the only adjustable parameters. 
Insets show the quality factor $Q$, defined as the number of oscillations before the amplitude damps by a factor of $e$, as a function of $J$ for both tilt and symmetric operation modes. Solid circles correspond to data in the main panel, and solid lines are theoretical predictions.
}
\label{fig:fig3}
\end{center}
\end{figure}

The oscillation frequency of $P_{\mathrm{S}}(\tau)$ gives a direct measure of $J$ at the exchange point X. Interestingly, the frequency does not depend on the Overhauser field, even when it is comparable in size to $J$~\cite{Barnes_next}. 
Figures~\ref{fig:fig3}(a) and (b) show a set of experimental exchange oscillations representative of the tilt and symmetric operation mode, respectively. 
$Q$ extracted from such oscillations is shown in the insets. 
Consistent with previous observations ~\cite{Petta_2005,Higginbotham_2014}, tilt-induced exchange oscillations result in $Q\sim6$ independent of $J$. On the other hand, for the symmetric mode, $Q$ increases with $J$ for the range measured of 40~MHz~$<J<$~700~MHz. This is in agreement with recent results in singlet-triplet qubits fabricated in the Si/SiGe heterostructures~\cite{Reed_2015}. Much higher values of Q can be obtained by tilting the double dot potential so far that both S and T$_0$ states share the same (0,2) charge state~\cite{Dial_2010}. However, it is unclear if qubit operations at frequencies of tens of GHz are practical.
 
To quantify the noise sensitivity of the symmetric exchange gate as well as gain insight into why it outperforms exchange by detuning, we compare both methods to a simple model that includes both nuclear Overhauser gradient noise and voltage noise on the detuning and barrier gates. Noise is assumed gaussian and quasistatic on the timescale of the exchange oscillations. Nuclear noise is characterized by a mean longitudinal Overhauser gradient energy $h_0$ between dots, with standard deviation $\sigma_h$. Exchange noise is assumed to result from voltage noise on left and right plungers and the barrier, with mean exchange energy $J$ with standard deviation $\sigma_J$.  The model also accounts for triplet-to-singlet relaxation at the measurement point, with a relaxation time $T_{\mathrm{RM}}$ during the measurement interval of length $T_\mathrm{M}$. Within this model, the singlet return probability $\langle\langle P_\mathrm{s}\rangle\rangle$ over both noise ensembles is given by~\cite{Barnes_next}:

 \begin{align}
	\label{Ed_Model}
	\begin{split}
	\langle\langle P_\mathrm{s}\rangle\rangle =&1-\frac{T_{\mathrm{RM}}}{T_\mathrm{M}}\left(1-e^{-\frac{T_\mathrm{M}}{T_{\mathrm{RM}}}}\right)
	\frac{e^{-\frac{h_0^2}{2\sigma_h^2}}e^{-\frac{J^2}{2\sigma_J^2}}}{\sqrt{\pi}\sigma_h\sigma_J}
	\\
	&
	\times\int_{-\pi/2}^{\pi/2} d\chi
	\Bigg\{\frac{b\left(\chi\right)}{a\left(\chi\right)^{3/2}} e^\frac{b\left(\chi\right)^2}{a\left(\chi\right)}
	\\
	&-\mathrm{Re}\Bigg[ \frac{b\left(\chi\right)+i\tau\mathrm{sec}(\chi)}{a\left(\chi\right)^{3/2}}
	e^\frac{[b\left(\chi\right)+i\tau\mathrm{sec}(\chi)]^2}{a\left(\chi\right)} \Bigg]\Bigg\}
	\end{split},
\end{align}
where $\chi$ is the tilt of the qubit rotation axis during an exchange pulse due to the Overhauser field gradient~\cite{Barnes_next}, $a\left(\chi\right) \equiv 2\mathrm{tan}^2\chi/\sigma_h^2+2/\sigma_J^2$ and $b\left(\chi\right)\equiv h_0\mathrm{tan}\chi/\sigma_h^2+J/\sigma_J^2$.

The black solid lines in Fig.~\ref{fig:fig3}, together with the insets in Figs.~\ref{fig:fig2}(a), (b) and (c), are generated by evaluating Eq.~\eqref{Ed_Model} numerically. Two fit parameters per curve are the oscillation frequency $J$ and a horizontal offset associated with the rise time of the waveform generator. All other parameters were obtained from independent measurements:
The Overhauser energy gradient fluctuations, $\sigma_h= 23$~MHz, was obtained by measuring the distribution of free induction decay frequencies \cite{Barthel_2012} over a 30 min.~interval and fitting the distribution to a gaussian.

The saturation of the singlet return probability, $P_S$, at long $\tau$, denoted $P_\mathrm{sat}$, will deviate from $P_\mathrm{sat}$~=~0.5 in the presence of a nonzero mean Overhauser field gradient, $h_0$, or finite relaxation time, $T_{\mathrm{RM}}$. Fitting the $J$ dependence of $P_\mathrm{sat}$ [Fig.~4(a)], yields fit values $T_{\mathrm{RM}}=30$~$ \mu$s and $h_0/h=40$~MHz. 

Exchange noise $\sigma_J$ is obtained by assuming (i) all noise is gate noise, (ii) noise on different gates is independent:
$\sigma_J^2=\sigma_\mathrm{el}^2[\left(dJ/dV_\mathrm{L}\right)^2+\left(dJ/dV_\mathrm{M}\right)^2+\left(dJ/dV_\mathrm{R}\right)^2]$.
In giving all three components equal weight, we have further assumed that all three gates are equally noisy as quantified by the parameter $\sigma_\mathrm{el}$. Taking into account the definitions in Eq.~\eqref{definition} we obtain:
\begin{align}
	\label{Ed_Model_complement}
	\begin{split}
	\sigma_J=\sigma_\mathrm{el}\sqrt{2k_0^2\left(\frac{dJ}{d\varepsilon_\mathrm{x}}\right)^2+\left(\frac{dJ}{d\gamma_\mathrm{x}}+k_1\frac{dJ}{d\varepsilon_\mathrm{x}}\right)^2}
	\end{split}
\end{align}

The derivatives are calculated from a phenomenological smooth exchange profile $J(\varepsilon_{\mathrm{x}},\gamma_{\mathrm{x}})$ fitted to a discrete map of $J$ measured at various operating points (see Supplementary Material~\cite{Supplement}). 
The effective gate noise $\sigma_\mathrm{el}$ is extracted from tilt exchange oscillations measured in a regime where effective detuning noise dominates, giving $\sigma_\mathrm{el}=0.18$~mV (see Supplementary Material). 
This value, together with Eq.~\eqref{Ed_Model_complement},  determines $\sigma_{J}(\varepsilon_{\mathrm{x}},\gamma_{\mathrm{x}})$ used in all simulations, and yields excellent agreement with data. 

\begin{figure}
\begin{center}
\includegraphics[width=86 mm]{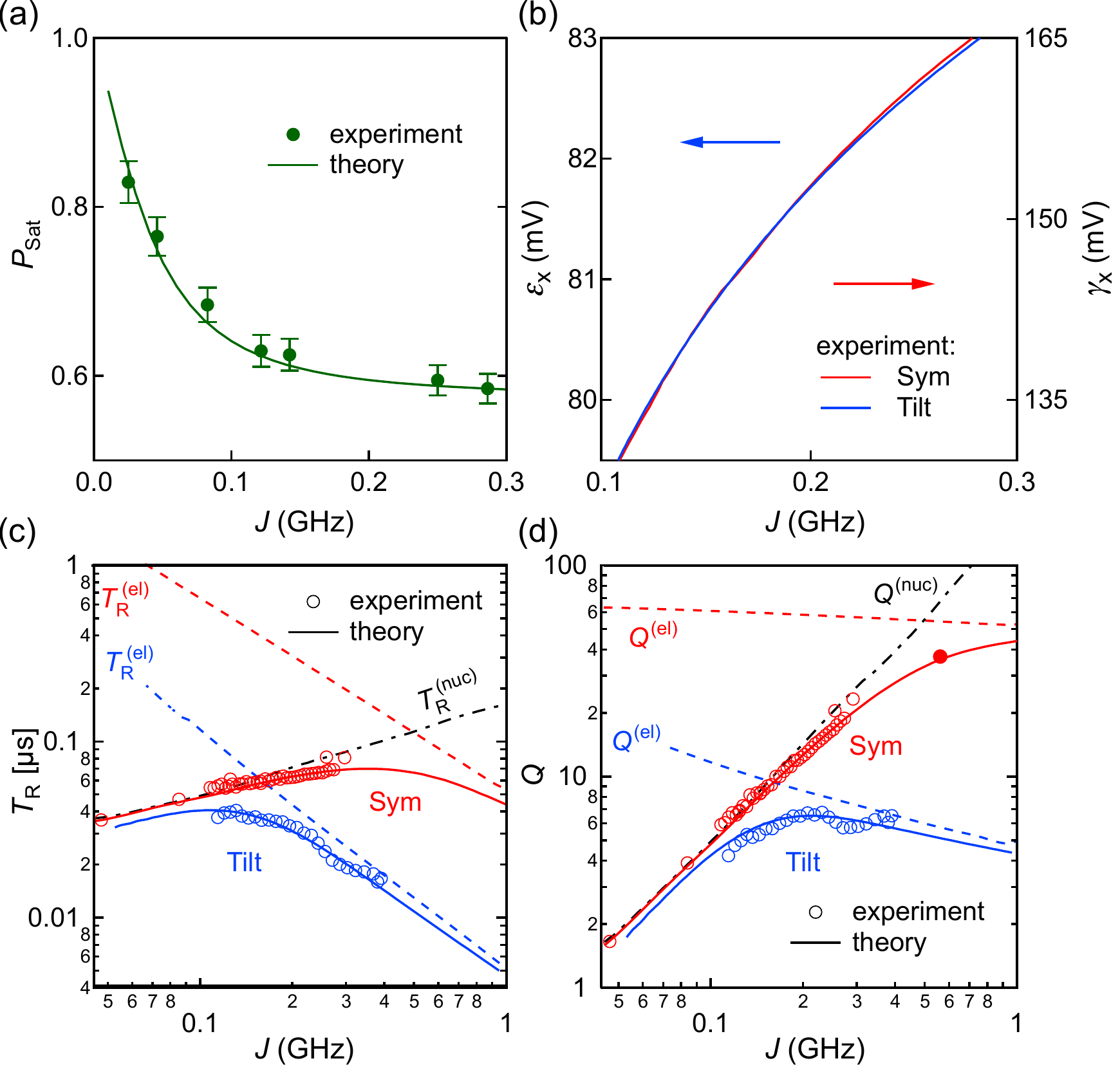}
\caption{(Color online)
(a) Saturation probability of the symmetric mode of operation, $P_{\mathrm{Sat}}$, as a function of $J$ (symbols). Comparison with theory (solid line) determines $T_\mathrm{RM}$ and $h_{\mathrm{0}}$.
(b) Plot of $\varepsilon_{\mathrm{x}}$ for the tilt  and  $\gamma_{\mathrm{x}}$ for the symmetric mode of operation, as functions of the exchange coupling extracted experimentally.
(c) Decoherence time $T_{\mathrm{R}}$, $i.e.$ time before the amplitude of oscillations is reduced by a factor of $e$, as a function of $J$ for both tilt and symmetric modes.
(d) Quality of the exchange rotations, defined as $Q = J T_{\mathrm{R}}$, for different $J$.
In (c) and (d) the open circles are obtained experimentally and solid lines correspond to a model that includes dephasing due to electrical and nuclear noise. 
Black dashed lines are the same model if we only consider nuclear noise contributions ($T_{\mathrm{R}}^{\mathrm{(nuc)}}$, $Q^{\mathrm{(nuc)}}$). Blue and red dashed lines correspond to the electrical noise contributions ($T_{\mathrm{R}}^{\mathrm{(el)}}$, $Q^{\mathrm{(el)}}$) for the tilt and symmetric modes of operation, respectively.
Solid circle indicates the maximum $Q$ value observed in Fig.~\ref{fig:fig2}(c).
}
\label{fig:fig4}
\end{center}
\end{figure}

The origin of the improved electrical performance becomes apparent when comparing the required pulse amplitudes for symmetric and tilted operation for a given $J$ [Fig.~\ref{fig:fig4}(b)].  Although the dependences of $\varepsilon_\mathrm{x}$ and $\gamma_\mathrm{x}$ on $J$ are  similar, the range of $\varepsilon_\mathrm{x}$ is significantly smaller than $\gamma_\mathrm{x}$. Note in Fig.~4(b) that $J$ changes from 0.1 to 0.3 GHz
for a $\sim$ 3 mV change in $\varepsilon_\mathrm{x}$, or a $\sim$ 30 mV change in $\gamma_\mathrm{x}$ [see Fig.~4(b)]. Because of this difference in derivatives of $J$ with respect to $\varepsilon_\mathrm{x}$ and $\gamma_\mathrm{x}$, the symmetric operation has much less noise for a given noise in the gate voltages.

The contributions of nuclear and electrical noise to limiting the quality factor $Q$ of and dephasing time, $T_\mathrm{R} = Q/J$, comparing experiment and model, is shown in Figs.~\ref{fig:fig4}(c) and (d).  
Note that for detuning (tilt) operation, electrical noise dominates above $\sim 0.2$ GHz, so that going any faster (using larger $J$) just makes the exchange noise greater in proportion, limiting the number of oscillations to $Q \sim 6$. For symmetric exchange, on the other hand, electrical noise doesn't dominate until above $J\sim 0.6$~GHz, resulting in a monotonically increasing quality factor up to $\sim 1$ GHz.
From the model, we find $Q$ as high as 50, 8 times larger than in the conventional tilt operation mode. 
Finally, we note that the origin of the effective electrical noise may be within the sample and not in the instrumentation. To distinguish actual voltage fluctuations on the gate electrodes (due to instrumentation) from intrinsic noise source (e.g. two-phonon processes~\cite{Kornich_2014}), further studies including temperature dependence are needed.      

In summary, we have investigated experimentally and modeled the application of an exchange gate applied by opening the middle barrier at a symmetry point of a two-electron spin qubit system instead of the conventional method, which is to detune the potential. 
The model allows the influences of nuclear and electrical noise to be disentangled for both symmetric and detuning exchange control, and is in excellent agreement with experimental data.
We find that symmetric mode of control is significantly less sensitive to electrical noise due to the symmetric arrangement, making exchange only quadratically sensitive to detuning gate voltage noise.  
With this new symmetric control method, we were able to increase the quality factor of coherent oscillations from around 6 to 35, and expect that improvements beyond $Q\sim 50$ are possible by further increasing $J$. 
The corresponding enhancement of coherence times by nearly an order of magnitude will also benefit other single- and multi-qubit implementations that rely on exchange interactions~\cite{Veldhorst_2015}. 

We thank Rasmus Eriksen for help in fast data acquisition and Daniel Loss and Mark S. Rudner for helpful discussions. 
This work was supported by IARPA-MQCO, LPS-MPO-CMTC, the EC FP7- ICT project SiSPIN no. 323841, the Army Research Office and the Danish National Research Foundation.

\clearpage
\onecolumngrid
\setcounter{figure}{0}
\setcounter{equation}{0}
\setcounter{page}{1}

\renewcommand{\thefigure}{S\arabic{figure}}  
\renewcommand{\theequation}{S\arabic{equation}}
\renewcommand{\thetable}{S\arabic{table}}
\renewcommand{\bibnumfmt}[1]{[S#1]}
\renewcommand{\citenumfont}[1]{S#1}

\section{Supplemental Material:}
\section{Noise suppression using symmetric exchange gates in spin qubits}

\vspace{0.1in}

\begin{enumerate}
\item \label{sec:epsilongamma}Relationship between control parameters $\varepsilon$, $\gamma$ and gate voltages $V_\mathrm{L}$,  $V_\mathrm{M}$, and $V_\mathrm{R}$
\item \label{sec:fittingJ}Extracting $J$, $T_\mathrm{R}$ and $Q$ from $P_\mathrm{s}(\tau)$
\item \label{sec:Jmodel}Model of $J(\varepsilon_{\mathrm{x}},\gamma_{\mathrm{x}})$
\item \label{sec:sigmaJmodel}Calculation of exchange noise $\sigma_J$, decoherence time $T_\mathrm{R}^\mathrm{(el)}$, and quality factor $Q^\mathrm{(el)}$ arising from quasistatic electrical noise $\sigma_\mathrm{el}$
\item \label{sec:extractsigmael}Determination of $\sigma_\mathrm{el}$
\item \label{sec:sigmaJcomparison}Comparison of electrical noise in tilt and symmetric operation
\item \label{sec:calcTQ}Calculation of $T_\mathrm{R}^\mathrm{(nuc)}$, $T_\mathrm{R}$, $Q^\mathrm{(nuc)}$ and $Q$.
\end{enumerate}

\vspace{0.2in}

\subsection{ \ref{sec:epsilongamma}. Relationship between control parameters $\varepsilon$, $\gamma$ and gate voltages $V_\mathrm{L}$,  $V_\mathrm{M}$, and $V_\mathrm{R}$}

Here, we explain in more detail the actual voltage pulses employed for qubit operation, and their relationship to control parameters for detuning and barrier height.

Fig.~S1 illustrates the pulse sequence.
Before each exchange pulse the qubit is prepared (P) in the eigenstate of the nuclear gradient field (denoted $|$$\uparrow\downarrow\rangle$) using standard voltage pulses applied to the left and right gate similar to earlier experiments~[S1]. These pulses realize exchange of electrons with the reservoirs to reset a singlet (0,2) state, a fast crossing of the $S$-$T_+$ degeneracy to avoid leakage into the $T_+$ state, and an adiabatic ramp to the (1,1) charge state that maps the singlet state into the $|$$\uparrow\downarrow\rangle$ state (I). After each exchange pulse the qubit is read out using standard voltage pulses applied to the left and right gate electrodes. These involve an adiabatic ramp and a fast crossing of the  $S$-$T_+$ degeneracy that maps the $|$$\uparrow\downarrow\rangle$ qubit state into a (0,2) singlet state or the $|$$\downarrow\uparrow\rangle$ qubit state into a (1,1) $T_0$ state. In the resulting measurement configuration (M), the charge states (0,2) and (1,1) are discriminated using single shot readout of the sensor quantum dot based on rf reflectometry and thresholding of the demodulated rf voltage~[S2, S3]. 

The exchange pulse itself differs from conventional operating schemes as it involves fast voltage pulses applied to left, middle and right gate electrodes ($V_{\mathrm{L}}$,$V_{\mathrm{M}}$ and $V_{\mathrm{R}}$ in Fig.~S1). For practical reasons, low-frequency and high-frequency signals are transmitted to the sample holder using twisted pairs and coax transmission lines, respectively, and combine on the sample holder using home-built RC bias tees.
After each readout pulse we apply pulse compensation pulses such that the time average of each coax voltage signal is equal to the voltage of the coax signal just before the exchange pulse. This means that the idling configuration of the qubit, characterized by voltages $V_{\mathrm{L}}^0$, $V_{\mathrm{M}}^0$, $V_{\mathrm{R}}^0$, corresponds to the DC voltages of the twisted pairs connected to the low-frequency input of the bias tees. This ensures that the idling configuration of the qubit does not change within a data set, even when changing amplitude or duration of the exchange pulses. 

In order to turn on a well-defined exchange splitting for a certain amount of time it is convenient to construct voltage pulses ($V_{\mathrm{L}}(t)-V_{\mathrm{L}}^0$, $V_{\mathrm{M}}(t)-V_{\mathrm{M}}^0$, $V_{\mathrm{R}}(t)-V_{\mathrm{R}}^0$) based on three control parameters $\delta$, $\varepsilon$, and $\gamma$:

\begin{figure}
\begin{center}
\includegraphics[width=120 mm]{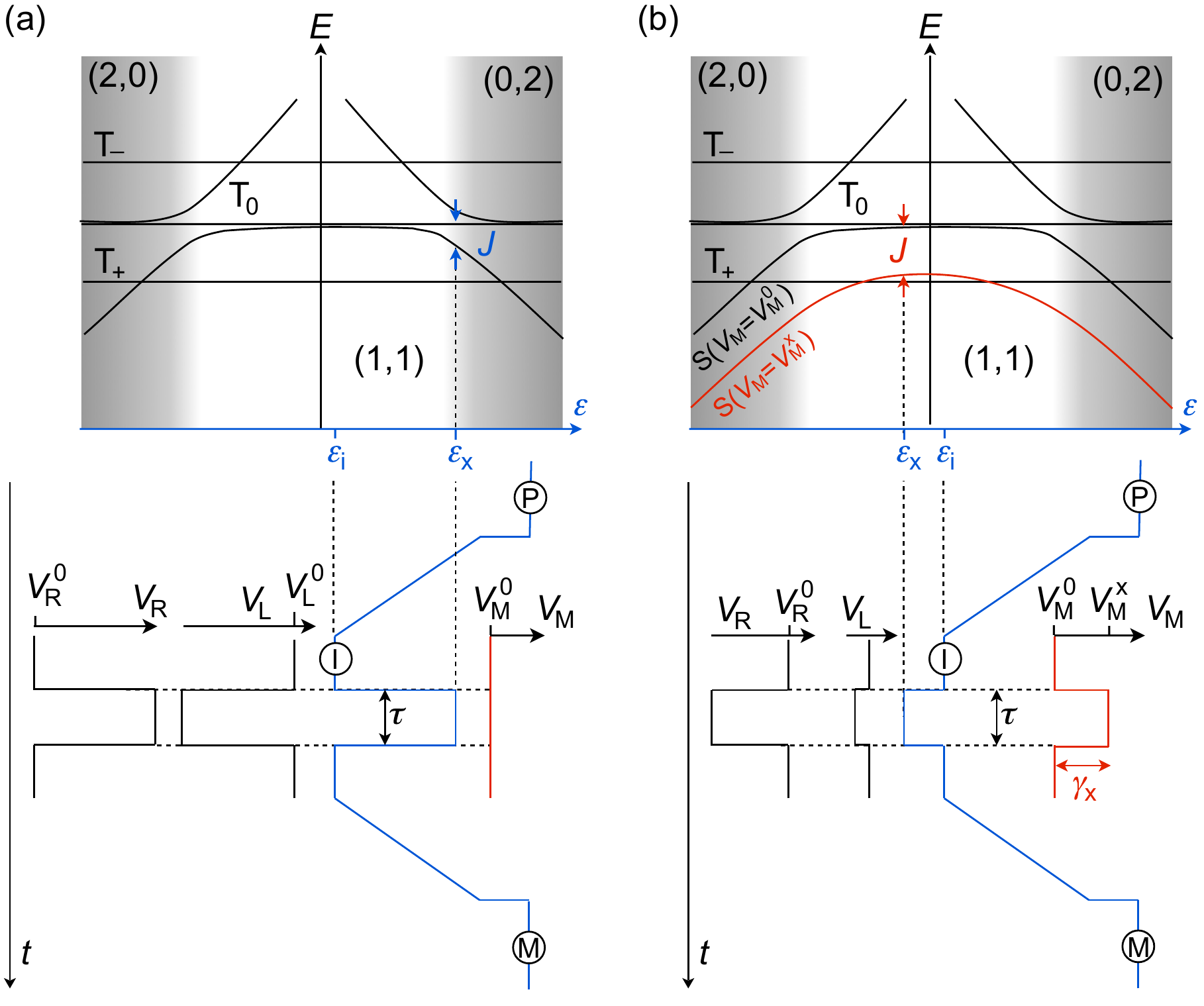}
\caption{(Color online) Schematic of the energy levels of the two-electron double quantum dot together with voltage pulses $V_{\mathrm{L}}(t)$, $V_{\mathrm{M}}(t)$, $V_{\mathrm{R}}(t)$ that implement a tilted (a) and barrier-induced exchange gate (b). The blue trace indicates detuning of the double dot during preparation of the singlet (P), after initialization of the $|$$\uparrow\downarrow\rangle$ state (I), and during the measurement of the charge sensor (M). 
}
\label{fig:figS1}
\end{center}
\end{figure}

\begin{figure}
\begin{center}
\includegraphics[width=160 mm]{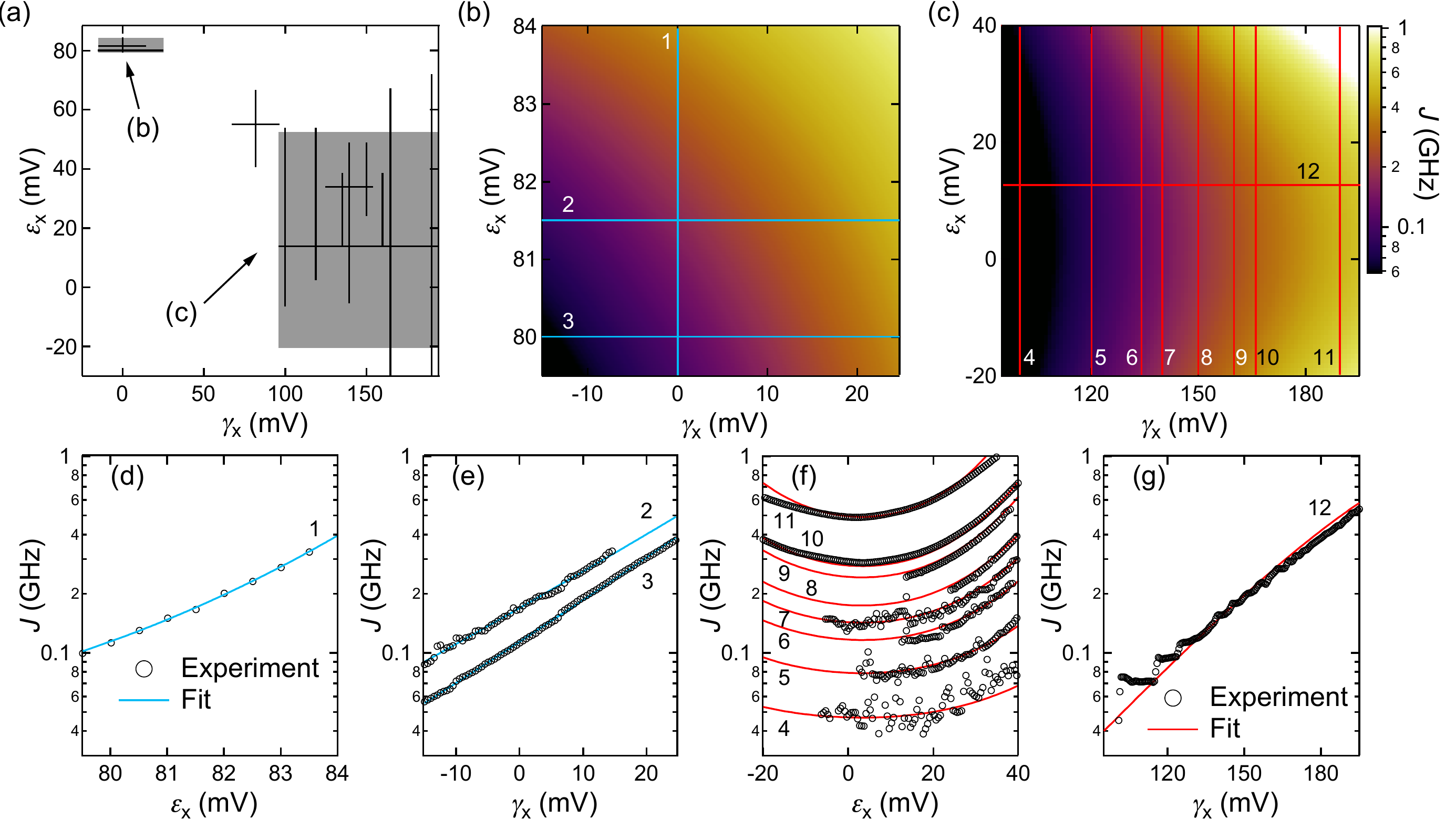}
\caption{(Color online)
(a) Range of control parameters $\varepsilon_{\mathrm{x}}$ and $\gamma_{\mathrm{x}}$ for which a numerical model of $J$ was developed (gray shaded regions). Each black line corresponds to operating points ($\varepsilon_{\mathrm{x}}$,$\gamma_{\mathrm{x}}$) where $P_\mathrm{s}(\tau)$ was measured. Extracting $J$ from $P_\mathrm{s}(\tau)$ on a subset of these lines yields the data points (symbols) plotted in panels (d-g). Fitting equation S2 to the symbols in panels (d-g) yields a two-dimensional model $J(\varepsilon_{\mathrm{x}},\gamma_{\mathrm{x}})$ presented in panels (b) and (c). Cuts indicated by numbers in (b) and (c) correspond to the solid lines in panels (d-g). 
}
\label{fig:figS2}
\end{center}
\end{figure}

\begin{align}
	\label{definition_0}
	 \begin{cases}
	 	V_{\mathrm{L}}-V_{\mathrm{L}}^0 & = \delta-\varepsilon-\alpha_1\gamma
		\\
		V_{\mathrm{R}}-V_{\mathrm{R}}^0 & = \delta+\varepsilon-\alpha_2\gamma
		\\
		V_{\mathrm{M}}-V_{\mathrm{M}}^0 & = \gamma
 	 \end{cases}
\end{align}
where, $\alpha_1 =0.675$ and $\alpha_2=0.525$.

These equations show that the three parameters $\delta$, $\varepsilon$, and $\gamma$ parameterize physically different manipulations of the (1,1) charge configuration. Namely, $\delta$ controls the common mode of the plunger gates (it appears with a plus sign in each equation) and brings the (1,1) charge state, which is in deep Coulomb blockade, toward the energy of the (2,2) or (0,0) charge states. In contrast, the detuning parameter $\varepsilon$ controls how much the double well potential is tilted towards the (0,2) charge state ($\varepsilon >0$) or the (2,0) charge state ($\varepsilon <0$). The barrier height in the double well potential is controlled by $\gamma$, which appears with a positive sign in the equation for $V_{\mathrm{M}}$ (i.e. positive $\gamma$ corresponds to lower barrier height/increased exchange splitting) and with a negative gain in $V_{\mathrm{L,R}}$ (in order to minimize its contribution to the common mode voltage).

For the symmetric operation of the exchange gate the choice of detuning and barrier during the time of exchange rotation, ($\varepsilon_{\mathrm{x}}$,$\gamma_{\mathrm{x}}$), are most important, as these parameters determine how much virtual tunneling to (0,2) and (2,0) can occur (setting the speed of the exchange gate), and how balanced these processes are (minimizing the sensitivity to $\varepsilon$ noise). The common mode voltage during the exchange pulse, $\delta_{\mathrm{x}}$, as well as the detuning voltage after initialization of the $|$$\uparrow\downarrow\rangle$ state, $\varepsilon_{\mathrm{i}}$, have a much weaker effect on the quality of observed exchange rotations, and therefore have not been studied systematically. For the measurement presented in the main text, we choose $\delta_{\mathrm{x}}=0$ and $\varepsilon_{\mathrm{i}}=13.5\mathrm{\ mV}$. 

\subsection{ \ref{sec:fittingJ}. Extracting $J$, $T_\mathrm{R}$ and $Q$ from $P_\mathrm{s}(\tau)$}

For each operating point of the exchange oscillation, $(\varepsilon_{\mathrm{x}},\gamma_{\mathrm{x}})$, the exchange interaction $J(\varepsilon_{\mathrm{x}},\gamma_{\mathrm{x}})$ can be determined by measuring $P_\mathrm{s}(\tau)$ and extracting the oscillation frequency. This method is justified even for $h_0\neq0$ provided that $P_\mathrm{s}(\tau)$ represents an average over a sufficiently large (quasistatic) ensemble characterized by $h_0<\sigma_h$. For our data sets, which typically involve averaging times exceeding 10 minutes, this condition is satisfied, and hence we do not have to take into account nuclear contributions to the oscillation frequency of type $\sqrt{J^2+h_0^2}$~[S4].

Specifically, we extract the frequency of the exchange oscillations for selected operating points [black lines in Fig.~S2(a)] in a two step process. First we calculate the discrete Fourier transform of $P_\mathrm{s}(\tau)$ and identify the main peak. Then we use the frequency associated with the main peak as an initial guess for fitting a damped sine wave of frequency $J$ to $P_\mathrm{s}(\tau)$, with a decay of the form exp$\left[-\left(\tau/T_\mathrm{R}\right)^\alpha\right]$. The quality factor is obtained using the relation $Q=J T_\mathrm{R}$. Values of $J$ obtained by this method are plotted as symbols in Fig.~S2(d-g).

\subsection{ \ref{sec:Jmodel}. Model of $J(\varepsilon_{\mathrm{x}},\gamma_{\mathrm{x}})$}

The operating points ($\varepsilon_{\mathrm{x}},\gamma_{\mathrm{x}}$) associated with data presented in panels Fig.~S2(d,e,f,g) fall onto a grid in the two-dimensional barrier-detuning space, as shown by black lines in  Fig.~S2(a). In order to inspect the sensitivity of $J$ to small fluctuations in  
$\varepsilon_{\mathrm{x}}$ and $\gamma_{\mathrm{x}}$ a two-dimensional model 
$J(\varepsilon_{\mathrm{x}},\gamma_{\mathrm{x}})$ is needed. 

Our phenomenological model of $J(\varepsilon_{\mathrm{x}},\gamma_{\mathrm{x}})$ is given by:

\begin{align}
	\label{fit}
 	J(\varepsilon_{\mathrm{x}},\gamma_{\mathrm{x}}) =e^{\left[c+\left(y_0+y_1\varepsilon_{\mathrm{x}}+y_2 \varepsilon_{\mathrm{x}}^2+y_3 \varepsilon_{\mathrm{x}}^3+y_4 \varepsilon_{\mathrm{x}}^4+y_5 \varepsilon_{\mathrm{x}}^6\right)\times\left((\gamma_{\mathrm{x}}-x_0) s_0+(\gamma_{\mathrm{x}}-x_1)^2 s_1\right)\right]} \mathrm{GHz}
\end{align}

\begin{table}[t]
\centering
\begin{tabular}{ |c|c|c| }
  \hline
 Parameters & Fig.~S2(b) & Fig.~S2(c)\\
 \hline  \hline  
$c$&  -2.62 &  -1.80\\
    \hline
$x_0$& 650  mV&-487  mV\\
  \hline 
$s_0$&0.351 mV$^{-1}$ &0.441 mV$^{-1}$\\
  \hline
$x_1$&-1695 mV& 1770 mV\\ 
  \hline
$s_1$  &8.01 $10^{-5}$ mV$^{-2}$&-7.21 $10^{-5}$ mV$^{-2}$\\
  \hline
$y_0$&0.205 & 0.0705 \\
    \hline
$y_1$& 8.39 $10^{-6}$ mV$^{-1}$& -9.39 $10^{-5}$ mV$^{-1}$\\
  \hline 
$y_2$& 2.11 $10^{-5}$  mV$^{-2}$ & 1.42 $10^{-5}$  mV$^{-2}$\\
  \hline
$y_3$& -1.67 $10^{-7}$  mV$^{-3}$ &5.80 $10^{-12}$ mV$^{-3}$\\ 
  \hline
$y_4$  & 8.25 $10^{-9}$ mV$^{-4}$ & 2.70 $10^{-9}$ mV$^{-4}$\\
  \hline
$y_5$ &-3.46 $10^{-13}$  mV$^{-6}$ & -3.80 $10^{-13}$ mV$^{-6}$\\
  \hline
\end{tabular}
\caption{Parameters for calculating the smooth exchange profile $J(\varepsilon_{\mathrm{x}},\gamma_{\mathrm{x}})$ from Eq. S2, shown in Figure~S2(b,c). These parameters were obtained by fitting Eq.~S2 to symbols in Fig.~S2(d-g).}
\label{tab:table1}
\end{table}

Using parameters from Table~\ref{tab:table1}, this model provides an excellent interpolation of $J$ in the gray shaded regions in Fig.~S2(a). 
These two regions have been selected based on the insight that they provide into the origin of the drastically different performance of tilted exchange gates and symmetric exchange gates (cf. section \ref{sec:sigmaJmodel} below).
Comparing line cuts of the model $J(\varepsilon_{\mathrm{x}},\gamma_{\mathrm{x}})$ with observed values of $J$ shows that our numerical model of $J$ accurately captures the observed exchange profile of the device [cuts of panels Fig.~S2(b,c) are shown as solid lines in panels Fig.~S2(d,e,f,g)]. The next section uses partial derivatives of this model to calculate the effects of effective gate noise.

\subsection{ \ref{sec:sigmaJmodel}. Calculation of exchange noise $\sigma_J$, decoherence time $T_\mathrm{R}^\mathrm{(el)}$, and quality factor $Q^\mathrm{(el)}$ arising from quasistatic electrical noise $\sigma_\mathrm{el}$}
In this section we describe how $T_\mathrm{R}^\mathrm{(el)}$, and quality factor $Q^\mathrm{(el)}$ in Fig.~4(c) and 4(d) in main text were calculated. 

We disregard nuclear fluctuations and consider decoherence caused by quasistatic effective gate noise only. Small fluctuations of control parameters $\varepsilon$ or $\gamma$ result in fluctuations of $J$ with an amplitude that is proportional to the partial derivative of $J$ with respect to $\varepsilon$ or $\gamma$. For comparison with other experiments, and in order to model the decoherence due to electrical noise, it is useful to express fluctuations of $J$ arising fo $\varepsilon$ or $\gamma$ noise in terms of effective gate noise on $V_\mathrm{L}$, $V_\mathrm{M}$, and $V_\mathrm{R}$.

From Eq.\ \ref{definition_0} we obtain the following relations between partial derivatives of $J$:

\begin{align}
	\label{definition_2}
	\begin{cases}
	\frac{dJ}{dV_\mathrm{L}} &= -k_{\mathrm{0}} \frac{dJ}{d\varepsilon}
	\\
	\frac{dJ}{dV_\mathrm{R}} &= k_{\mathrm{0}} \frac{dJ}{d\varepsilon}
	\\
	\frac{dJ}{dV_\mathrm{M}} &= k_{\mathrm{1}} \frac{dJ}{d\varepsilon}+\frac{dJ}{d\gamma}
	\end{cases}
\end{align}

Assuming that the effective gate noise associated with $V_{\mathrm{L}}$, $V_{\mathrm{M}}$ and $V_{\mathrm{R}}$ is quasistatic, independent, and Gaussian distributed with a common standard deviation $\sigma_\mathrm{el}$, i.e. $\sigma_\mathrm{L}=\sigma_\mathrm{M}=\sigma_\mathrm{R}=\sigma_\mathrm{el}$, we can write the expected fluctuations of $J$:

\begin{align}
	\label{EdEquationSupp}
	\begin{split}
	\sigma_J=\sqrt{\left(\frac{dJ}{dV_\mathrm{L}}\sigma_L\right)^2+\left(\frac{dJ}{dV_\mathrm{M}}\sigma_M\right)^2+\left(\frac{dJ}{dV_\mathrm{R}}\sigma_R\right)^2}=\sigma_\mathrm{el}\sqrt{2k_0^2\left(\frac{dJ}{d\varepsilon}\right)^2+\left(\frac{dJ}{d\gamma}+k_1\frac{dJ}{d\varepsilon}\right)^2} 
	\end{split}
\end{align}

Averaging over a quasistatic, Gaussian ensemble of $J$ with standard deviation $\sigma_J$ yields a Gaussian decay envelope $\exp\left[-(\tau/T_{\mathrm{R}})^2\right]$, a decoherence time given by $T_{\mathrm{R}}^\mathrm{(el)}=1 /\left(\sqrt{2}\pi\sigma_J\right)$, and a quality factor given by $Q^\mathrm{(el)}= JT_{\mathrm{R}}^\mathrm{(el)}$~[S4, S5]. To generate the associated curves in Fig.~4(c) and 4(d) in the main text we use $\sigma_\mathrm{el}$~=~0.18~mV, determined as described in the next section.

\subsection{ \ref{sec:extractsigmael}. Determination of $\sigma_\mathrm{el}$}

We determine the effective gate noise $\sigma_\mathrm{el}$ by measuring tilt-induced exchange oscillations in a regime where effective detuning noise $\sigma_\varepsilon$ dominates, i.e. for $\varepsilon_{\mathrm{x}}$~=~84~mV. 
Fitting a sinusoid with a Gaussian envelope, $ \exp\left[-(\tau/T_{\mathrm{R}})^2\right]$, yields $T_{\mathrm{R}}= 14$ ns. Using $T_{\mathrm{R}}=1/\left(\sqrt{2}\pi\sigma_J\right)$ yields $\sigma_J~=~1.6$ MHz. 
Taking in account Equations~S2, S3 and S4 and Table~S1 this value corresponds to an effective gate noise $\sigma_\mathrm{el}$~=~0.18~mV.

\subsection{ \ref{sec:sigmaJcomparison}. Comparison of electrical noise in tilt and symmetric operation}

Application of Eq.S4 to the model $J(\varepsilon_{\mathrm{x}},\gamma_{\mathrm{x}})$ shown in Fig.~S2(b) and (c) allows us to calculate  $\sigma_J$ [Fig.~S3(a) and (b)].
To highlight the difference in magnitude of $\sigma_J$ between tilt and symmetric mode of operation we use the same color scale for panels (a) and (b), and compare two cuts plotted against $J$ in panel (c). 
This analysis demonstrates that the same amount of effective gate noise ($\sigma_\mathrm{el}$~=~0.18~mV) results in exchange noise ($\sigma_J$) that is more than one order of magnitude larger in the tilt mode of operation than the symmetric mode of operation, for  a given $J$.

\subsection{ \ref{sec:calcTQ}. Calculation of $T_\mathrm{R}^\mathrm{(nuc)}$, $Q^\mathrm{(nuc)}$}

Here we describe how theoretical curves $T_\mathrm{R}^\mathrm{(nuc)}$ and $Q^\mathrm{(nuc)}$ in Fig. 4(c,d) of the main text were calculated. 
These quantities represent the contribution of nuclear noise to the total noise.
First, exchange oscillations were simulated using Eqs. 3 of the main text for both tilt and symmetric mode of operations, similar to simulating insets shown in Fig. 2(a) and (b) in the main text, but using $\sigma_\mathrm{el}=0$ while keeping all other parameters unchanged. 
From these simulations, $T_\mathrm{R}^\mathrm{(nuc)}$ and $Q^\mathrm{(nuc)}$ were extracted in the same way as $T_\mathrm{R}$ and $Q$ were extracted as described in section~\ref{sec:fittingJ}.

\begin{figure}
\begin{center}
\includegraphics[width=160 mm]{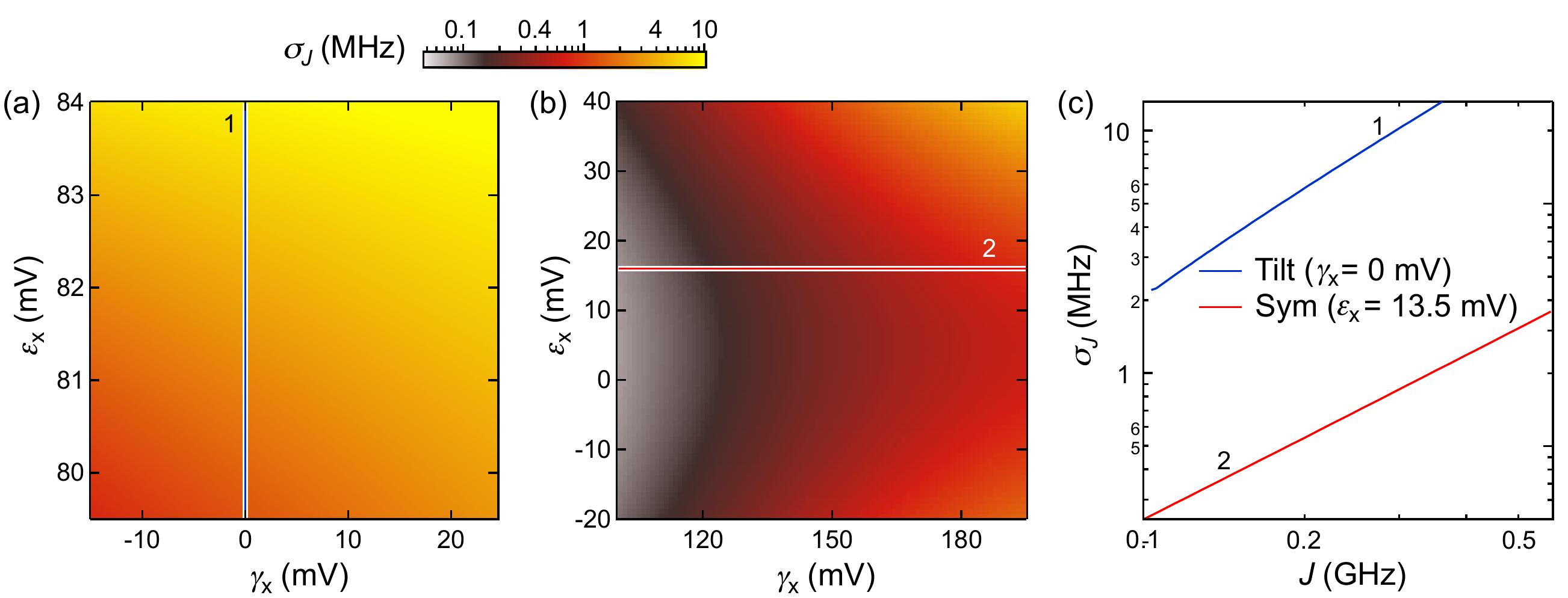}
\caption{(Color online)
(a) and (b) $J$ noise $\sigma_J$ as a function of $\varepsilon$ and $\gamma$ for the surfaces in Fig.~\ref{fig:fig2}(b) and (c). Both panels assume the same amount of effective gate noise ($\sigma_\mathrm{el}$~=~0.18~mV).
(c) $\sigma_J$, extracted from cuts 1 and 2 indicated in panel (a) and (b), as a function of $J$ for tilt and symmetric mode of operation.
}
\label{fig:figS3}
\end{center}
\end{figure}


\begin{thebibliography}{10}
\bibitem{Kloeffel_2013}
C.~Kloeffel and .~D.~Loss, Ann.~.~Rev.~Condens.~Matter Phys.~ {\bf 4}, 10.1-10.31 (2013).
\bibitem{Taylor_2005}
J.~M.~Taylor, H.-A.~Engel, W.~D\"ur, A.~Yacoby, C.~M.~Marcus, P.~Zoller, and M.~D.~Lukin, Nat.~Phys.~{\bf 1}, 177 (2005).
\bibitem{Awschalom_2013}
D.~D.~Awschalom, L.~C.~Bassett, A. ~S.~Dzurak, E.~L.~Hu, J.~R.~Petta, Science {\bf339}, 1174 (2013).
\bibitem{FolettiNatPhys2009}
S.~Foletti, H.~Bluhm, D.~Mahalu, V.~Umansky, and A.~Yacoby, Nature Phys.~{\bf5}, 903 (2009). 
\bibitem{Bluhm_2010_1}
H.~Bluhm, S.~Foletti, D.~Mahalu, V. Umansky, A. Yacoby, Phys. Rev Lett. {\bf 105}, 216803 (2010).
\bibitem{Petta_2010}
J. R. Petta, H. Lu, A. C. Gossard, Science {\bf 327}, 670 (2010).
\bibitem{ShulmanScience2012}
M. D. Shulman, O. E. Dial, S. P. Harvey, H. Bluhm, V. Umansky, and A .Yacoby, Science {\bf 336} 202 (2012).
\bibitem{Dial_2010}
O. E. Dial, M. D. Shulman, S. P. Harvey, H. Bluhm, V. Umansky, and A. Yacoby, Phys. Rev. Lett. {\bf110}, 146804 (2013).
\bibitem{Barthel_2012}
C. Barthel, J. Medford, H. Bluhm, A. Yacoby, C. M. Marcus, M. P. Hanson, and A. C. Gossard, Phys. Rev. B {\bf 85}, 035306 (2012).
\bibitem{Bluhm_2010_2}
H. Bluhm, S. Foletti, I. Neder, M. Rudner, D. Mahalu, V. Umansky, and A. Yacoby, Nat. Phys. {\bf7}, 109 (2010).
\bibitem{WongPRB2015}
C. H. Wong, M. A. Eriksson, S. N. Coppersmith, and M. Friesen, Phys. Rev. B {\bf 92}, 045403 (2015).
\bibitem{HiltunenPRB2015}
T. Hiltunen, H. Bluhm, S. Mehl, and A. Harju, Phys. Rev. B {\bf 91}, 075301 (2015).
\bibitem{FeiPRB2015}
J. Fei, J-T Hung, T. S. Koh, Y-P Shim, S. N. Coppersmith, X. Hu, and M. Friesen, Phys. Rev. B {\bf 91}, 205434 (2015).
\bibitem{DobrovitskiPRL2010}
V. V. Dobrovitski, G. de Lange, D. Rist\`e, and R. Hanson, Phys. Rev. Lett {\bf 105}, 077601 (2010).
\bibitem{CerfontainePRL2014}
P. Cerfontaine, T. Botzem, D. P. Divincenzo, and H. Bluhm, Phys. Rev. Lett. {\bf 113}, 150501 (2014).
\bibitem{Loss_1998}
D. Loss and D. P. DiVincenzo, Phys. Rev. A {\bf57}, 120 (1998).
\bibitem{Petta_2005}
J. R. Petta, A. C. Johnson, J. M. Taylor, E. A. Laird, A. Yacoby, M.D. Lukin, C.M. Marcus, M.P. Hanson, and A. C. Gossard, Science {\bf309}, 2180 (2005).
\bibitem{Bertrand_2014}
B. Bertrand, H. Flentje, S. Takada, M. Yamamoto, S. Tarucha, A. Ludwig, A. D. Wieck, C. Bauerle and T. Meunier, Phys. Rev. Lett.  {\bf115}, 096801 (2015).
\bibitem{Reed_2015}
M. D. Reed, B. M. Maune, R. W. Andrews, M. G. Borselli, K. Eng, M. P. Jura, A. A. Kiselev, T. D. Ladd, S. T. Merkel, I. Milosavljevic, E. J. Pritchett, M. T. Rakher, R. S. Ross, A. E. Schmitz, A. Smith, J. A. Wright, M. F. Gyure, A. T. Hunter, arXiv:1508.01223 (2015).
\bibitem{Barthel_2009}
C. Barthel, D. J. Reilly, C. M. Marcus, M. P. Hanson, and A. C. Gossard,  Phys. Rev. Lett. {\bf103}, 160503 (2009).
\bibitem{Buizert_2008}
C. Buizert, F. H. L. Koppens, M. Pioro-Ladri\`ere, H-P Tranitz, I. T. Vink, S. Tarucha, W. Wegscheider, and L. M. K. Vandersypen, Phys. Rev. Lett. {\bf101}, 226603 (2008).
\bibitem{Barthel_2010}
C. Barthel, M. Kjaergaard, J. Medford, M. Stopa, C. M. Marcus, M. P. Hanson, and A. C. Gossard, Phys. Rev. B {\bf81}, 161308R (2010).
\bibitem{definitionsymm}
Fig.~\ref{fig:fig2}(c) shows that barrier-induced exchange oscillations have high quality factors for a wide range of operating points $\varepsilon_{\mathrm{x}}$ near the sweet spot. For example, data shown in Fig.~\ref{fig:fig2}(b) was obtained at $\varepsilon_{\mathrm{x}}=13.5$ mV, which for practical purposes we also classify as symmetric operation. 
\bibitem{Barnes_next}
E. Barnes, M. S. Rudner, F. Martins, F. K. Malinowski, C. M. Marcus and F. Kuemmeth, arXiv:1511.07362 (2016).
\bibitem{Supplement}See Supplemental Material, which includes Ref.~\cite{Reilly_2007_2}.
\bibitem{Reilly_2007_2} 
D. J. Reilly, C. M. Marcus, M. P. Hanson, and A. C. Gossard, App. Phys. Lett. {\bf91}, 162101 (2007).
\bibitem{Higginbotham_2014}
A.~P. Higginbotham,  F.~Kuemmeth, M. P. Hanson, A.~C. Gossard, and C.~M. Marcus, Phys. Rev. Lett. {\bf112}, 026801 (2014).
\bibitem{Kornich_2014}
V. Kornich, C. Kloeffel, and D. Loss, Phys. Rev. B {\bf89}, 085410 (2014).
\bibitem{Veldhorst_2015}
M. Veldhorst, C. H. Yang, J. C. C. Hwang, W. Huang, J. P. Dehollain, J. T. Muhonen, S. Simmons, A. Laucht, F. E. Hudson, K. M. Itoh, A. Morello, and A. S. Dzurak, Nature {\bf526}, 7573 (2015).

\end{thebibliography}

\begin{thebibliography}{8}%
\makeatletter
\providecommand \@ifxundefined [1]{%
 \@ifx{#1\undefined}
}%
\providecommand \@ifnum [1]{%
 \ifnum #1\expandafter \@firstoftwo
 \else \expandafter \@secondoftwo
 \fi
}%
\providecommand \@ifx [1]{%
 \ifx #1\expandafter \@firstoftwo
 \else \expandafter \@secondoftwo
 \fi
}%
\providecommand \natexlab [1]{#1}%
\providecommand \enquote  [1]{``#1''}%
\providecommand \bibnamefont  [1]{#1}%
\providecommand \bibfnamefont [1]{#1}%
\providecommand \citenamefont [1]{#1}%
\providecommand \href@noop [0]{\@secondoftwo}%
\providecommand \href [0]{\begingroup \@sanitize@url \@href}%
\providecommand \@href[1]{\@@startlink{#1}\@@href}%
\providecommand \@@href[1]{\endgroup#1\@@endlink}%
\providecommand \@sanitize@url [0]{\catcode `\\12\catcode `\$12\catcode
  `\&12\catcode `\#12\catcode `\^12\catcode `\_12\catcode `\%12\relax}%
\providecommand \@@startlink[1]{}%
\providecommand \@@endlink[0]{}%
\providecommand \url  [0]{\begingroup\@sanitize@url \@url }%
\providecommand \@url [1]{\endgroup\@href {#1}{\urlprefix }}%
\providecommand \urlprefix  [0]{URL }%
\providecommand \Eprint [0]{\href }%
\providecommand \doibase [0]{http://dx.doi.org/}%
\providecommand \selectlanguage [0]{\@gobble}%
\providecommand \bibinfo  [0]{\@secondoftwo}%
\providecommand \bibfield  [0]{\@secondoftwo}%
\providecommand \translation [1]{[#1]}%
\providecommand \BibitemOpen [0]{}%
\providecommand \bibitemStop [0]{}%
\providecommand \bibitemNoStop [0]{.\EOS\space}%
\providecommand \EOS [0]{\spacefactor3000\relax}%
\providecommand \BibitemShut  [1]{\csname bibitem#1\endcsname}%
\let\auto@bib@innerbib\@empty
 \bibitem{Petta_2005}
J. R. Petta, A. C. Johnson, J. M. Taylor, E. A. Laird, A. Yacoby, M.D. Lukin, C.M. Marcus, M.P. Hanson, and A. C. Gossard, Science {\bf309}, 2180 (2005).
\bibitem{Reilly_2007} 
D. J. Reilly, C. M. Marcus, M. P. Hanson, and A. C. Gossard, App. Phys. Lett. {\bf91}, 162101 (2007).
\bibitem{Barthel_2009} 
C. Barthel, D. J. Reilly, C. M. Marcus, M. P. Hanson, and A. C. Gossard,  Phys. Rev. Lett. {\bf103}, 160503 (2009).
\bibitem{Barnes_next}
E. Barnes, M. S. Rudner, F. Martins, F. K. Malinowski, C. M. Marcus and F. Kuemmeth, arXiv:1511.07362 (2016).
\bibitem{Dail_2010}
O. E. Dial, M. D. Shulman, S. P. Harvey, H. Bluhm, V. Umansky, and A. Yacoby, Phys. Rev. Lett. {\bf110}, 146804 (2013).
\end{thebibliography}
\end{document}